# Generation Alpha: Understanding the Next Cohort of University Students

Rushan Ziatdinov [a,*], Juanee Cilliers [b]

[a] Department of Industrial Engineering, Keimyung University, 704-701 Daegu, South Korea
[b] Faculty of Design, Architecture and Building, University of Technology Sydney, 2230 Sydney, Australia

**Abstract.** Technology is changing at a blistering pace and is impacting on the way we consider knowledge as a free commodity, along with the ability to apply skills, concepts and understandings. Technology is aiding the way the world is evolving, and its contributions to education are not an exemption. While technology advances will play a crucial part in future teaching-learning approaches, educators will also be challenged by the next higher-education generation, the Alpha Generation. This entrepreneurial generation will embrace the innovation, progressiveness, and advancement with the expectation that one in two Generation Alphas will obtain a university degree. In anticipating the educational challenges and opportunities of the future higher education environment, this research reflected on Generation Alpha as the next cohort of university students, considering their preferred learning styles, perceptions and expectations relating to education. The research employed a theoretical analysis based on the characteristics and traits that distinguishes Generation Alpha, spearheaded by technology advances. The empirical investigation considered three independent studies that were previous conducted by authors from Slovakia, Hungary, Australia, and Turkey to understand the challenges and opportunities pertaining to Generation Alpha. The research identified the influence of social media, social connections, high levels of perceptions and the Generation Alpha's ability to interpret information as strengths to consider in future teaching-learning approaches in the higher education environment. This research concluded with recommendations on how universities could be transformed to ensure a better learning experience for Generation Alpha students, aligned with their characteristics, perceptions and expectations.

---

[*] Corresponding author.
E-mail addresses: ziatdinov@kmu.ac.kr, ziatdinov.rushan@gmail.com (Rushan Ziatdinov), jua.cilliers@uts.edu.au (J. Cilliers).
URLs: https://www.ziatdinov-lab.com/ (Rushan Ziatdinov), https://profiles.uts.edu.au/Jua.Cilliers (J. Cilliers).

### 1. Introduction to Generation Alpha

*Generation Alpha* is composed of individuals who were born at the crossover of *Generation Z* and the new age (Tootell et al., 2014). This is the new generation that will soon fill classrooms and universities and demand unique approaches to teaching-learning, based on their unique skillsets and requirements.

According to the model proposed by Howe and Strauss (1991), a generation change occurs approximately every 20 years, with certain signs of cyclicality. *Generation Y*, composed of people born in the 1980s and the 1990s, were described with the label "*MTV Generation*" mainly due to the influence and relevance of the music channel during their time. Meanwhile, there is Generation Z, the first-ever generation to have technology and social media as a vast part of their day-to-day lives.

As it is known, Generation Alpha, the succeeding generation, are no strangers and soon to be frontiers of this highly digitized world. Technology is aiding the way the world is evolving, and its contributions to education are not an exemption. The recent COVID-19 pandemic outbreak resulted in a global uptake in distanced learning, which was forcefully implemented to continue education amidst lockdown restrictions and stay-at-home orders. In an attempt to foster social distancing and slow the virus's spread (Viner et al., 2020), students and parents turned to mobile devices such as smartphones, tablets and laptops to access classroom information (Nadeak, 2020). Parents had to take on a role in guiding their children, particularly Generation Alpha, through their education during the distance learning set-up, more than they usually did during face-to-face classes. In one particular study, parents had mixed responses regarding the struggles imposed upon by distance learning. Some of them included balancing responsibilities, learner needs, personal balance, lack of motivation both related and unrelated to remote learning, accessibility, learner content needs, lack of pedagogy, lack of connectivity and resources, and need for teacher communication among others (Garbe et al., 2020). The COVID-19 pandemic accelerated the debate on online teaching and related methodologies, and while academic institutions and universities are currently exploring delivery options, better student engagement and ways to enhance the student experience, it is clear that technology is heavily infused in the culture and environment of Generation Alpha, and it will just as greatly be incorporated into their education going forward. This paper aims to reflect on the differences between Generation Alpha and previous generations, particularly in regards to their education on the tertiary level pertaining to the teaching-learning experience, their perceptions and expectations.

## 2. The anatomy of Generation Alpha

Mark McCrindle, an Australian social researcher, futurist, and demographer, was the first to propose the term "*Generation Alpha*". The name marks newness and not a return or stay in the old. Generation Alpha is regarded as the twenty-first century's second true generation. It births date from 2010 and beyond, which implies most of these students are currently still in their school years (Amrit, 2020). According to Amrit (2020), Generation Alpha students have from a very young age, exposure to the marketing, technology, traveling, and priorities of their millennial parents. Ironically, the year that marks the birth of this generation is the same year that the word "app" has been declared the word of the year (Amrit, 2020). These youth have more access to technology, information, and external influences than any previous generation (McCrindle & Fell, 2020) and as a result this generation will be highly captured through app-based play, more screen time, shorter attention spans, and a lack of digital literacy combined with a lack of social formation (McCrindle & Fell, 2020).

Generation Alpha is quite different from the preceding generations, especially because their reality, and all aspects of life, has been dominated by technology. Generation Alpha is growing up in unprecedented times of change and rapid technology innovation, and they are part of an inadvertent worldwide experiment in which screens have been placed in front of them as pacifiers, entertainment, and educational aides since they were very small. (McCrindle & Fell, 2020). Generation Alpha has been labeled as generation glass, screenagers, digital natives, and the connected or wired generation because of their clear connection to technology and technological innovation (Tootell et al., 2014). Them being born in a highly digitalized world ultimately gives them an advantage, in engaging with the primary machineries that we have today, found in the form of hand-held gadgets such as smartphones, iPads, laptops, and the like. Naturally, they are faster and more in-depth when it comes to learning about what technology has to offer as opposed to the previous generations. Through the day-to-day engagement they have with their gadgets, they can learn, even on their own, making them well-versed in interacting and living in this post-modern world. Generation Alpha is unlikely to carry a wallet, use single-use plastics, listen to the radio as a device, take a written exam, or set an analogue alarm clock (McCrindle & Fell, 2020).

The parents of Generation Alpha are being more aware of both advantages and disadvantages of early exposure of their children to technology (McCrindle & Fell, 2020). While these parents embrace the benefits of the technology advances, they are likewise aware of the skills that the Generation Alpha student would need in future, especially pertaining to social competencies, entrepreneurial skills, strength and coordination, financial literacy, innovation, and resourcefulness and (McCrindle & Fell, 2020).

## 3. Reflection of studies conducted on Generation Alpha

Ziatdinov, R. & Cilliers, J. (2021). Generation Alpha: Understanding the Next Cohort of University Students, European Journal of Contemporary Education 10(3): 783-789.

This section provides a reflection of previous conducted studies by various authors to understand the challenges and opportunities pertaining to Generation Alpha in relation to teaching-learning in the higher education environment.

### 3.1. Research piece 1: Generation Alpha, Marketing or Science?

The study by Slovakian and Hungarian researchers Nagy and Kölcsey (2017) used traditional desk research to create a generation paradigm. This study also mentioned the apparent similarities that Generation Alpha holds and shares with Generation Z. Included in the similarities is the fact that social media platforms have more influences on them, along with the vast changes in their learning styles, resulting in the need for innovative teaching methods. This is largely due to the fact that they are constantly flooded with information and have quicker access to it. Generation Alpha's criticisms were clear in their dislike of the sharing economy (shouting "Mine!" and "All mine!" and refusing to share anything). Generation Alpha appears to be unconcerned about privacy and rules, not confined to boundaries, and to live in the now. The study concludes that the label given to this generation is based on marketing rather than science. The study determined that Gen Alpha is similar to its predecessors, but that it carries on their "legacy" (Nagy & Kölcsey, 2017), despite the fact that further research is needed to fully comprehend this new generation.

### 3.2. Research piece 2: A Generation Alpha case study

Research by Australian researchers Taylor and Hattingh (2019) critiqued and dissected the Four Resource Model (FRM) reading practices in playing Minecraft, a serial video game, as applied and implemented by Generation Alpha. The skills of code breaker, text participant, text user, and text analyzer were all considered in the FRM. Observations, field notes, semi-structured interviews, and a researcher reflective journal were used to support the research (Taylor & Hattingh, 2019) in terms of language and articulation; social and mentor integration; real-world linkage; and parent and child perspectives. The study identified that children were able to apply reading skills within Minecraft, even children with foundational reading skills were able to use words repetitively and interpret information (Taylor & Hattingh, 2019). It was also highlighted that the social aspect and option to interact with other players were received positively by the children. It was evident that the children were thoroughly engaged and ardent when it came to playing the game. The study's conclusion supports the ways that children, particularly Generation Alpha, learn through technology like Minecraft. In the study posed valuable information about learning about the ways Generation Alpha prefers to learn, especially in theorizing the way their tertiary education will be shaped.

### 3.3. Research piece 3: Preschool Teachers' Views on Generation Alpha

Ziatdinov, R. & Cilliers, J. (2021). Generation Alpha: Understanding the Next Cohort of University Students, European Journal of Contemporary Education 10(3): 783-789.

The study of researchers from Turkey, Apaydin & Kaya (2020) investigated pre-school teachers' perceptions of Generation Alpha pertaining to the classroom setup and learning process. The research adopted a qualitative design and included the inputs from teachers of private kindergartens in Antalya from 2018-2019. The research acknowledged the digital environment which Generation Alpha has inhibited upon their birth and noticed the technology literacy that most teachers and educators lack, pondering over its possible impact on the quality of education that will be delivered and served to Generation Alpha. The research highlighted some negative characteristics of Generation Alpha including, technology addiction, the tendency to be egocentric, and the tendency to violence (Apaydin & Kaya, 2020). High levels of perception, tapping out with music, effective use of numbers, being meticulous, and emotional were all positive characteristics of Generation Alpha. In terms of comparing Generation Z and Generation Alpha it was evident that Generation Alpha are more open towards knowledge and the generality of things, have high numerical intelligence, but limited social intelligence. Both generations share the similarity of a tendency towards technology (Apaydin & Kaya, 2020). According to the findings, Generation Alpha expects visual, aural, and kinesthetic methods to be used in classroom management, and they are more prone to distractions, which are crucial factors to take into consideration in creating a teaching-learning environment.

## 4. Discussion

### 4.1. Future collegiate circumstances and expectations for Generation Alpha

Generation Alpha's education will be primarily influenced by technological advancements (Romero, 2017). Born from parents from the Millennial Generation, they are more tech-savvy, more entrepreneurial, and willing to create their own jobs (Romero, 2017). Their career choices and life decisions will also differ from those taken by former generations, simply because of the innovation, progressiveness, and advancement that is largely pervasive in the world they are currently living in. The amount that they are likely to conform to prejudices, biases, and norms established by society will significantly be less. Their formal education has never been equaled in the history of the world, with a predicted one in two Generation Alphas to obtain a university degree (McCrindle & Fell, 2020). Digital skills combined with creativity, curiosity, and adaptability can be expected as Generation Alpha's fortes and core competencies. On the other hand, Generation Alpha shall work upon improving their critical thinking skills and leadership skills (McCrindle & Fell, 2020).

### 4.2. Implications for higher education

As the world continuously undergoes rapid shifts brought about by post-modernity, education is also advancing and adapting technology within its curriculum models. In the present day, academic institutions are already accepting the need for better integration of technology into education. Long a hallmark of academic study, technological innovation may now be transforming the way institutions educate and students learn (Glenn, 2008). Universities are embracing transformational benefits such as distance education, advanced learning management systems, and the ability to work with research partners from all over the world (Glenn, 2008). Various other studies confirmed that the inclusion of technology increases the learning and interactivity of students, and that modern students prefer to use technology for educational support (Raja & Nagasubramani, 2018). Interactivity, ease, convenience, and accessibility are prevalent factors that are steadily defining good education. In this perspective, technology has four roles in the sphere of education: 1) it is part of the curriculum, 2) it is used as an educational delivery system, 3) it is used to aid instructions, and 4) it is used to enhance the entire learning process, allowing education to be interactive rather than passive (Raja & Nagasubramani, 2018). The crucial challenges in this regard relates to teachers who might not be as technologically adept as the Alpha Generation (Prensky, 2001), and that changes in the current structures and mindset of institutions are usually slow, as is adoption of new pedagogical approaches (Romero, 2017). As such, the culture prevalent in academia will be greatly challenged. Universities would need to further diversify, not merely for social justice, but because students need to refer to more to people who can represent them (Romero, 2017). In prioritizing the student-centered and community-based learning model, experiential learning would need to be included in mainstream teaching-learning, enabling students to reflect on the learning process and even learning from failed experiments (Romero, 2017).

Experiential learning focus on learning-by-doing and the experiences gained through reflection on doing. It requires the student to take initiative, to make decisions, and be accountable for the outcomes. It is built upon actions of investigating, experimenting, problem solving, accountability, creativeness, and the integration previously developed through the process of doing (Itin, 1999).

New approaches to teaching, such as experiential learning, would need to be considered, approaches that will work for students who are vastly different from the typical in terms of culture, education and expectations (Romero, 2017). Universities would need to develop the soft skills that would be crucial in the modern world, including critical thinking, problem-solving, teamwork and communication abilities (Romero, 2017). Table 1 provides an analysis of the research findings and the associated implications for teaching that would need to be considered in future educational approaches.

Table 1: Analysis of research findings and implications for teaching.

| Research | Findings | Implications for teaching |
| --- | --- | --- |
| Nagy and Kölcsey (2017) | Social media have direct influence | Vast changes in learning styles needs to be incorporated in teaching-learning approaches. Social media's impact on learning effectiveness and student experience needs to be recognized. |
| | Quick access to information | Focus should be on knowledge development, not only accessing information. Interpretation of information is crucial. |
| | Detests sharing economy | Soft skill development should be prioritized sharing and public goods as shared commodity. |
| | Not confined to boundaries | Out-of-the box approaches and experiential learning will be essential for future learning activities. |
| Taylor and Hattingh (2019) | Apply reading skills online | Traditional learning methods can be subconsciously developed through online gaming portals. It is about keeping the students' interest and attention. |
| | Ability to interpret information | Translation of information to knowledge is essential in teaching the next generation. |
| | Highlight social connections | Social connections are possible in the online and virtual environment, it would need better planning and coordination to create a good student experience. |
| | Learn through technology | Technology is a tool for enhanced education. Through technology, |
| | | teaching outcomes can be achieved, but it would require a unique design, interactive portal and continuous support base. |
| Apaydin and Kaya (2020) | Lack of technology literacy | The technology literacy gap between teachers and students are prominent and identified as the greatest challenge for online learning. |
| | High levels of perception | Students show high levels of perception that is developed from their interaction with technology from a young age. This should be considered an advantage to teaching. |
| | Visual, auditory and kinesthetic tools | Visual, auditory and kinesthetic tools will characterize the future teaching-learning environment in attempt to mimic the technology advances that are 'normal' to the Generation Alpha student within the educational space. |

Generation Alpha is a young generation and the body of research dealing with Generation Alpha is still relatively limited.

While technology is replacing jobs, it is also creating a slew of new ones, as evidenced by the current Fourth Industrial Revolution (McCrindle & Fell, 2020). Many students of today are garnering and honing skills in big data analytics, robotics, social media marketing, and app development (McCrindle & Fell, 2020). These skills will be crucial for jobs that are still yet to exist in the very near future, which will be saturated by today's learners and those to come. These jobs will both consider the changes witnessed by technology and demography. Careers in new industries such as cyber-security, software development, and cryptocurrencies will be available to Generation Alphas (McCrindle & Fell, 2020). They will be tenured in handling several jobs at once, continuously learning throughout their lifetime. They will also need to be adaptable, regularly upskilling and retraining to stay current with the changes they will face as they progress through their careers (McCrindle & Fell, 2020). The role of universities would be to harness these skills and prepare Generation Alpha students to embrace the digital world they know so well, to optimize skills and experience to co-create the solutions that our future will need.

## 5. Conclusions and recommendations

The Generation Alpha students' learning style will be largely dependent and connected to technology. Technology advances will likewise have an impact on their learning effectiveness and the overall student experience. Experiential learning will play a key part of the future teaching-learning approaches, especially to engage students and to enable them to co-create knowledge, and not just merely access information instantly. It will be about the translation of information, the interpretation information and adding of value. Visual, auditory and kinesthetic tools will support the future teaching-learning environment, to provide a real experience with supporting social connections. In this sense, the challenge would be to bridge the literacy gap between teachers and students to enhance the social connections and interactions, and to develop soft skills that will foster a sense of belonging, of community and of sharing. The lecturer stands central to creating a collaborative, critical-thinking, and co-creative classroom atmosphere (Steyn, 2015). This demands an educator with a robust academic point of departure; who is well-versed and educated; able to develop knowledge and transfer core disciplinary principles to a new generation of students; someone who understands educational theories as commons; and the role of the university within broader society and the knowledge community. The instructor must think critically and imaginatively in order to create a classroom environment that is conducive to thinking and creating (Steyn, 2015) that is based on the Generation Alpha student's perceptions and expectations.

Higher education will in future most likely involve technology-integrated learning programs and options, far more career engaging and career preparation events, and scarce skill development training and programs. The focus will shift from 'transfer of knowledge' to 'co-creation of knowledge', optimizing the skillset of the Generation Alpha student and their unique acceptance and understanding of technology advances.